\documentclass[acmsmall,nonacm]{acmart}

\AtBeginDocument{%
  \providecommand\BibTeX{{%
    \normalfont B\kern-0.5em{\scshape i\kern-0.25em b}\kern-0.8em\TeX}}}

\usepackage{changes} 
\usepackage{algorithm}
\usepackage{algpseudocode}
\usepackage[utf8]{inputenc}
\usepackage[T1]{fontenc}
\usepackage{textcomp} % 可能需要这个包来处理某些特殊字符
\usepackage{amsmath}
\usepackage{booktabs}
\usepackage{multirow}
\usepackage{siunitx}
\usepackage{tikz}
\usepackage{paralist}
\usepackage{pgfplots}
\pgfplotsset{compat=1.17}
\usepackage{xspace}
\usepackage{url}
\usepackage{array}
\usepackage{adjustbox}
\usepackage{float}
\usepackage{tabularx}
\usepackage{enumitem}
\usepackage{threeparttable}
\usepackage{graphicx}
\usepackage{lineno}
\usepackage{subcaption}
\usepackage{svg}
\begin{document}
\title{Compiler Bugs Detection in Logic Synthesis Tools via Linear Upper Confidence Bound}

\author{Hui Zeng}
\email{zh974757632@dlmu.edu.cn}
\affiliation{%
  \institution{Dalian Maritime University}
  \country{China}
}

% \author{Shikai Guo}
% % \thanks{Corresponding author: Shikai Guo (Email: shikai.guo@dlmu.edu.cn)}
% \email{shikai.guo@dlmu.edu.cn}
% % \authornote{Corresponding authors.}
% \affiliation{%
%   \institution{Dalian Maritime University}
%   \country{China}
% }

\author{Zhihao Xu}
\email{Zhihao.Xu@monash.edu}
\affiliation{%
  \institution{Monash University}
  \country{Australia}
}

% \author{Wen Zhao}
% \email{zhaowen2003@dlmu.edu.cn}
% \affiliation{%
%   \institution{Dalian Maritime University}
%   \country{China}
% }

\author{Hui~Li}
\email{li_hui@dlmu.edu.cn}
\affiliation{%
  \institution{Dalian Maritime University}
   \country{China}
}

\author{Siwen~Wang}
\email{wsw@dlmu.edu.cn}
\affiliation{
  \institution{Dalian Maritime University}
   \country{China}
}

\author{Qian~Ma}
\email{maqian@dlmu.edu.cn}
\affiliation{
  \institution{Dalian Maritime University}
   \country{China}
}

%%
%% The abstract is a short summary of the work to be presented in the
%% article.

\begin{abstract}

Field-Programmable Gate Arrays (FPGAs) play an indispensable role in Electronic Design Automation (EDA), translating Register-Transfer Level (RTL) designs into gate-level netlists. The correctness and reliability of FPGA logic synthesis tools are therefore critically important, as unnoticed bugs in these compilers can propagate into final hardware implementations, potentially leading to severe safety and security issues.
Recent methods have advanced the testing of FPGA logic synthesis tools by systematically generating Hardware Description Language (HDL) test cases. However, these approaches often rely heavily on random selection strategies, limiting the structural diversity of the generated HDL test cases and resulting in inadequate exploration of the tool’s feature space.
To address this limitation, we propose Lin-Hunter, a novel testing framework designed to systematically enhance both the diversity of HDL test cases and the efficiency of FPGA logic synthesis tool validation. 
Specifically, Lin-Hunter introduces a principled set of metamorphic transformation rules to generate functionally equivalent yet structurally diverse HDL test case variants, effectively addressing the limited diversity of existing test inputs. 
To further enhance bug discovery efficiency, Lin-Hunter integrates an adaptive strategy selection mechanism based on the Linear Upper Confidence Bound (LinUCB) method. 
This method leverages feedback from synthesis logs of previously executed test cases to dynamically prioritize transformation strategies that have empirically demonstrated a higher likelihood of triggering synthesis bugs.
Comprehensive experiments conducted over a three-month period demonstrate the practical effectiveness of Lin-Hunter. 
Our method has  discovered 18 unique bugs, including 10 previously unreported defects, which have been confirmed by official developers. 
Our results highlight the capability of  Lin-Hunter in efficiently uncovering critical bugs. 
And our method has demonstrated outperforming state-of-the-art testing methods in both test-case diversity and bug-discovery efficiency.
\end{abstract}


\maketitle
\section{Introduction}

Field-Programmable Gate Arrays (FPGAs) hold an irreplaceable significance in modern electronic design~\cite{andina2017fpgas,monterrosa2016design}. FPGAs are highly flexible and programmable integrated circuits that can be configured and reconfigured post-manufacturing using Hardware Description Languages (HDLs).~\cite{sharma2010fpga,chiuchisan2013new,dillinger2005fpga} This makes FPGAs suitable for fields with high demands for safety and stability, such as aerospace, automotive electronics, and medical devices.~\cite{shreejith2014extensible,sanaullah2018real,chiuchisan2013new,sklyarov2011fpga,varga2015c,anish2013ethernet,tietche2012fpga}

\begin{figure}[!t]
  \centering
  \includegraphics[width=\linewidth]{ 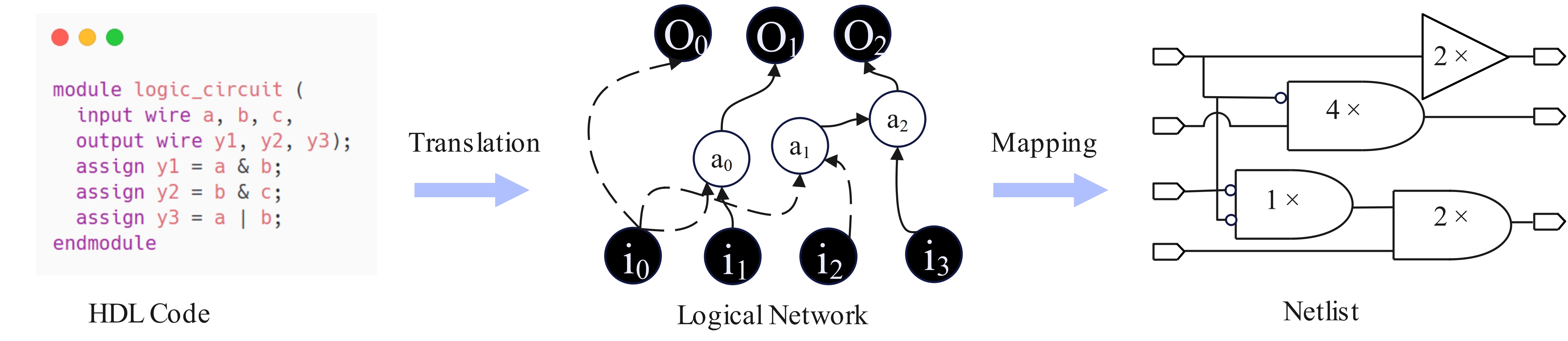}
  \caption{Electronic Circuit Design - Logic Synthesis Flowchart}
  \label{fig:aig}
\end{figure}

Logic synthesis plays a fundamental and indispensable role in FPGA design and development. Specifically, logic synthesis tools allow engineers to systematically transform Register-Transfer Level (RTL) descriptions into gate-level netlists, serving as a critical bridge between software specifications and hardware implementations. As illustrated in \figurename~\ref{fig:aig}, logic synthesis can translate high-level software descriptions into corresponding hardware netlists, facilitating their deployment on FPGA devices~\cite{singh1995extracting}.

With the increasing demand for intelligent and automated hardware design, ensuring the correctness of FPGA logic synthesis tools has become increasingly crucial. Bugs within logic synthesis compilers may propagate silently into the final hardware, potentially causing device malfunctions, severe safety hazards, or other critical issues. Moreover, mis-translations arising during logic synthesis transformations may result in unintended or incorrect hardware behaviors. Unfortunately, these errors are often overlooked by engineers or mistakenly attributed to hardware-level faults. Thus, the rigorous validation and testing methodologies for logic synthesis tools are becoming more and more important.

Several methods have been proposed to generate HDL code for testing FPGA logic synthesis tools, typically using fuzzing test methods to randomly generate HDL test cases. 
To the best of our knowledge, Verismith~\cite{Verismith} is the most widely used method for testing FPGA logic synthesis tools. 
It uses an Abstract Syntax Tree (AST)-based generation method to create HDL test cases.
However, Verismith has some limitations, such as high test case redundancy and the ability of generating single hardware description language. 
These limitations hinder deeply testing of FPGA logic synthesis tools. 
% Furthermore, redundant test cases lead to require more time to find bugs, as Verismith has been  
% so that Verismith only finds 11 bugs in two-year period. 
Therefore, LegoHDL~~\cite{LegoHDL} was proposed to address the limitations of high test case redundancy and the constraint of generating only one hardware description language. This approach transforms the task of HDL code generation into the generation of Cyber-Physical Systems (CPS) models and utilizes HDL Coder to convert these CPS models into HDL code for FPGA logic synthesis testing. With comprehensive support for the CPS module library, LegoHDL can generate more complex test cases than Verismith~\cite{LegoHDL}.

As both Verismith and LegoHDL are built upon fuzz testing, they inevitably inherit its intrinsic stochastic behavior, which hinders further testing of logic synthesis tools~\cite{fuzz}.
Therefore, it may also result in a lack of diversity in the generated HDL test cases. Specifically, there are two main challenges in testing FPGA logic synthesis tools.

\textbf{Challenge 1. The diversity of HDL test cases.}
The randomness of the CPS models guiding the generation of HDL code leads to insufficient diversity in the HDL code generated based on this model.
For example, by examining the CPS source code, we found that LegoHDL generates a large number of repetitive mathematical modules when creating CPS models. 
This results in similar mathematical operations in the HDL code. Therefore, although the test cases generated by LegoHDL are more complex compared to Verismith, it still should be improved. 
So, how to generate diverse HDL test cases that can further explore the entire testing space is the first challenge in testing FPGA logic synthesis tools.

\textbf{Challenge 2. The efficiency of bug discovery.}
Low bug discovery effectiveness remains a significant limitation in the testing of FPGA logic synthesis tools. Despite advances in fuzzing-based techniques, such as those employed by Verismith, the rate of bug identification remains relatively low. For example, Verismith identified only 11 unique bugs over a span of two years, consuming approximately 16,000 CPU hours in the process~\cite{Verismith}. This translates to an average of around 1,450 CPU hours per confirmed bug, underscoring the inefficiency inherent in current approaches. Several factors contribute to this limitation, including the vast input space of Verilog programs, the high semantic redundancy among generated test cases, and the absence of precise guidance toward triggering deep semantic failures. As a result, enhancing the efficiency of bug discovery—both in terms of time and computational cost—emerges as a critical challenge in the domain of automated testing for FPGA logic synthesis tools.
% An increasing number of HDL test cases with repetitive or similar data flows and control flows could hinder comprehensive testing in FPGA synthesis tools. 
% This limitation makes it difficult to thoroughly explore the entire testing space, causing developers to spend more time identifying and fixing bugs.

% We have summarized the overall testing cycles of existing FPGA logic synthesis testing methods and calculated the average bugs discovery time to evaluate the bugs discovery efficiency.
% We used two metrics to measure their bug discovery efficiency: the time required to find a single bug and the number of test cases needed to find a single bug.
% As shown in Figures \ref{fig:subfigA} and \ref{fig:subfigB}, the average time required to find a single bug over than 200 CPU hours and the number of test cases needed to find over than 500000 HDL code files.
% For instance, Verismith discovered 11 FPGA logic synthesis tool bugs over a period of 2 years (16000 CPU hours), which was written in their paper. 
% On average, Verismith takes approximately 1450 hours to discover one bug. 
% When generating code using the configuration recommended by Verismith (700-1000 lines of code) on an i9@2.1GHz CPU, Verismith needs to generate around 500,000 test cases to find one bug.
% Therefore, how to improve the efficiency of bug discovery is the second challenge in testing FPGA logic synthesis tools.

To address the limitations, we proposed a novel method \textbf{Lin-Hunter}, which employs metamorphic relationship construction to diversify HDL test cases while utilizing the Linear Upper Confidence Bound (Lin\_UCB) algorithm to guide the metamorphic relationship strategy selection and enhance bug discovery efficiency in FPGA logic synthesis tools.
Lin-Hunter includes three main components; the Metamorphic relationship construction component, the Metamorphic strategy selection component, and the differential testing component. 
The key insight of Lin-Hunter is to efficiently construct diverse HDL test cases with metamorphic relationships to thoroughly test FPGA logic synthesis tools.

Specifically, to address the challenge of insufficient diversity in generated HDL test cases, the Metamorphic Relationship Construction component introduces a set of four well-defined transformation rules designed to produce functionally equivalent yet structurally diverse variants. These metamorphic transformation rules systematically modify both the control-flow and data-flow structures of the original test cases, without altering their intended functional semantics. By doing so, this component significantly enriches the syntactic and structural variety of test inputs, thereby improving the likelihood of exposing hidden bugs in logic synthesis tools that may be sensitive to specific code patterns or structural features.

% Subsequently, to enhance the testing efficiency of FPGA logic synthesis tools and maximize the discovery of bugs, the Metamorphic strategy selection component employs the linear upper confidence bound to guide the metamorphic relationship strategy selection. 
% By using the bug information Lin-Hunter can update the reward results based on the synthesis log information, thereby selecting mutation strategies that are more likely to trigger bugs.

% Finally, the differential testing component compares the output consistency between equivalent but structurally different HDL test case variants to detect bugs in FPGA logic synthesis tools.

Subsequently, to improve the testing efficiency of FPGA logic synthesis tools and to maximize bug discovery, the Metamorphic Strategy Selection component adopts a linear upper confidence bound (LinUCB)-based decision-making algorithm. This approach dynamically guides the selection of metamorphic transformation strategies by balancing exploration (testing less-used strategies) and exploitation (prioritizing historically effective ones). In particular, Lin-Hunter leverages synthesis-time feedback—such as anomaly patterns and log messages—to assign adaptive rewards to different strategies, updating its selection policy based on prior bug-triggering performance. This guided approach significantly reduces redundant test generation and focuses computational resources on transformations more likely to reveal previously undetected bugs.

Finally, the Differential Testing component serves as the bug detection backbone of the framework. It compares the synthesis results of HDL test case variants that are semantically equivalent but structurally different, as generated through metamorphic transformations. Any inconsistency in the outputs—such as mismatched netlists, divergent resource utilization, or synthesis failures—is flagged as a potential indicator of a logic synthesis bug. This differential comparison approach is particularly effective in exposing subtle defects that may not manifest under uniform test structures, thereby enhancing both the depth and breadth of tool validation.

% To evaluate the effectiveness of Lin-Hunter, we conduct comprehensive experiments including bug detection and the comparison with SOTA methods.
% Over the course of three months, Lin-Hunter detected 18 bugs, 10 of which were newly discovered and all of which were confirmed by the official developers. 
% Additionally, we conducted ablation experiments on the key components of Lin-Hunter to validate our approach. 
% The results demonstrated that using the Lin\_UCB algorithm to guide CPS model mutations selection is four times more effective at bug detection compared to random mutation strategies and twice as effective as mutation strategies using the $\epsilon$-greedy method.

To assess the effectiveness of Lin-Hunter, we conducted a comprehensive set of experiments focusing on both bug detection capability and comparative performance against state-of-the-art (SOTA) methods. Over a testing period of three months, Lin-Hunter successfully identified 18 unique bugs, among which 10 were previously undiscovered. All reported bugs were subsequently acknowledged and confirmed by the official developers of the respective FPGA logic synthesis tools, underscoring the practical value and reliability of our approach.
In addition, we performed a series of ablation studies to evaluate the contribution of key components within the Lin-Hunter framework. The results demonstrated that employing the LinUCB algorithm to guide CPS (control and path structure) mutation strategy selection significantly enhances bug detection efficiency. Specifically, Lin-Hunter achieved a 4× improvement in bug discovery rate compared to purely random mutation, and a 2× improvement over the commonly used $\epsilon$-greedy exploration strategy. 

The main contributions of our work are as follows:
\begin{itemize} 
\item We propose Lin-Hunter, a novel testing framework for FPGA logic synthesis tools. Lin-Hunter introduces a metamorphic testing approach that leverages the semantic equivalence of structurally diverse mutations to generate high-diversity HDL test cases. Furthermore, it incorporates a LinUCB-based adaptive strategy selection mechanism, which dynamically updates the reward model based on synthesis log feedback. This allows Lin-Hunter to prioritize mutation strategies with higher bug-triggering potential, thereby improving testing efficiency.

\item We conduct extensive experiments to evaluate the effectiveness of Lin-Hunter in both HDL test generation and bug detection. Results show that Lin-Hunter can uncover bugs significantly faster and with greater diversity compared to state-of-the-art methods. Over a three-month period, Lin-Hunter discovered 18 unique bugs, 10 of which were previously unknown and later confirmed by the official tool developers, demonstrating its practical effectiveness and impact..

\item To support reproducibility and further research, we have made our code publicly available on GitHub~~\cite{Lin-Hunter}, facilitating future investigations.
\end{itemize}

The remainder of this paper is organized as follows. Our motivation is discussed in Section 2. The main components of our proposed model are introduced in Section 3. The experimental setup and results are provided in
Section 4. Related work is discussed in Section 5. Section 6 summarizes our work and outlines future directions.

\begin{figure}[t]
\centering
\includegraphics[width=\linewidth]{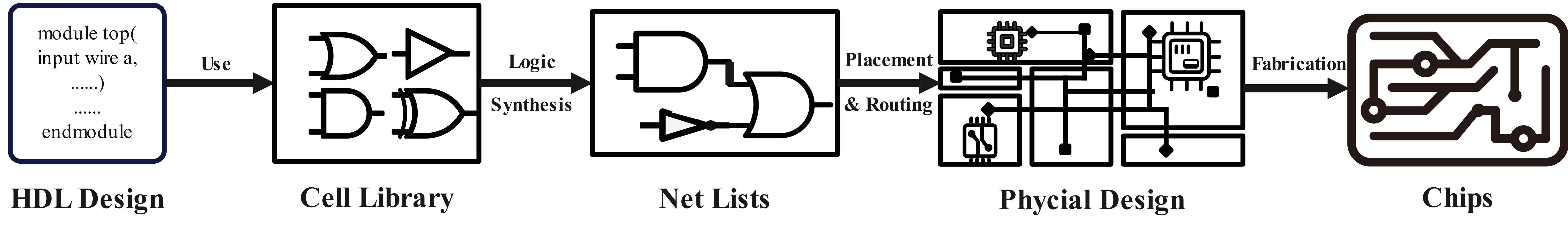}
\caption{FPGA design Flowchart}
\label{fig:figure1}
\end{figure}

\section{Background and Related Work}
\label{sec:related}
\subsection{FPGA Logic Synthesis Workflow}
Modern FPGA development follows a standardized design process (illustrated in Figure~\ref{fig:figure1}), comprising four core phases: Hardware Description Language (HDL) design, logic synthesis, simulation verification, and physical implementation. Engineers first construct Register-Transfer Level (RTL) circuit models using Verilog HDL, defining digital circuit architectures and dataflow characteristics through modular design principles~\cite{ashenden2007digital}. Subsequently, the logic synthesis tool converts the RTL code into a gate-level netlist composed of fundamental logic components such as Look-Up Tables (LUTs), Flip-Flops (FFs), and others, while performing critical technological operations. These operations include technology library mapping, timing constraint parsing, and application of optimization strategies, aimed at enhancing circuit performance, area efficiency, and power consumption effectiveness\cite{singh1995extracting}.

To ensure design correctness, a multi-dimensional verification framework is essential: functional verification employs testbenches to apply stimulus signals, combining formal methods with dynamic simulation; timing analysis establishes clock tree models for setup/hold time verification and critical path optimization~\cite{bryant1991methodology}. The final implementation phase maps netlists to target FPGA devices through place-and-route tools, completing clock domain partitioning, I/O constraint configuration, and power optimization~\cite{abboud2008mathematical,sharma2005accelerating}.

As the critical bridge between software and hardware, the quality of logic synthesis tools directly determines chip performance. However, their validation faces dual challenges: synthesis errors are frequently misattributed to design flaws, while the analog nature of hardware signals complicates fault tracing. These factors necessitate the development of robust automated testing frameworks.

\subsection{Related Work}

\subsubsection{\textbf{Bug Detection in Logic Synthesis Tools}}
Current testing approaches for FPGA synthesis tools exhibit three generations of technical evolution:

First-generation random testing frameworks, exemplified by VlogHammer\footnote{\url{https://yosyshq.net/yosys/vloghammer.html}}, employ syntax-driven random code generation. While effective for basic test case creation, their limited expressiveness prevents modeling multi-module interaction scenarios and behavioral-level Verilog constructs.

Second-generation AST-based methods achieved breakthroughs through Verismith~\cite{Verismith}. This tool generates structurally valid Verilog code via syntax tree mutations, incorporating differential testing to compare outputs across synthesis tools. Over two years of testing, it detected 11 tool bugs. However, corpus limitations constrain code diversity, hindering deep compiler vulnerability detection.

Third-generation intelligent generation methods demonstrate diversified development. VeriGen~\cite{verigen} utilizes CodeGen-16B-based LLM techniques, enhancing semantic compliance through fine-tuning on open-source Verilog corpora. However, its generated test cases diverge significantly from real-world engineering scenarios, lacking targeted stimulus construction capabilities. LegoHDL~\cite{LegoHDL} innovates by reformulating hardware modeling as Cyber-Physical System (CPS) block composition, automatically generating HDL code through model transformation. While detecting 20 tool bugs within three months, its random block assembly strategy causes rigid module interfaces and excessive code redundancy.

Existing methods confront two persistent challenges: 1) The structural complexity gap between test cases and real-world designs limits activation of deep state machine errors; 2) Corpus bugs combined with parametric randomness create combinatorial explosions, severely reducing high-value test case generation efficiency.

\subsubsection{\textbf{Evolution of Differential Testing Techniques}}
Differential testing detects behavioral discrepancies through semantic-equivalent variants, demonstrating unique advantages in compiler verification. The Ccoft framework~\cite{tu2022detecting} employs structured syntax mutations to construct differential testing scenarios for C++ compiler frontends, uncovering 136 bugs in GCC/Clang within three months with 135\% efficiency improvement. Similarly, RustSmith~\cite{sharma2023rustsmith} generates type-safe Rust programs through ownership model constraints, identifying memory management vulnerabilities in rustc via cross-compiler validation.

In EDA testing, the integration of differential testing with formal methods has spawned next-generation verification frameworks. SLFORGE~\cite{slforge} pioneered hybrid verification architecture combining stochastic model generation with basic differential testing, though its mutation operators were limited by static dead code elimination. SLEMI~\cite{slemi} revolutionized this approach through runtime feature analysis and equivalence modulo input (EMI) techniques, effectively identifying zombie code via path-sensitive mutation strategies. Tran et al.~\cite{Tran} further integrated model refactoring with formal verification, enabling semantics-preserving complex transformations through compositional mutation operators, albeit requiring model checkers for behavioral consistency.

Notably, adaptive improvements of traditional mutation testing in EDA have emerged. Zhan et al.~\cite{Zhan} proposed dynamic taint analysis-based path-sensitive mutation criteria, significantly reducing equivalent mutant generation by enforcing mutation effect propagation along all feasible paths. This methodological evolution reflects the paradigm shift from random exploration to targeted bug mining in hardware verification.

\begin{figure}[t]
\centering
%\includesvg[width=\linewidth]{figs/framework3.svg}
\includegraphics[width=0.9\linewidth]{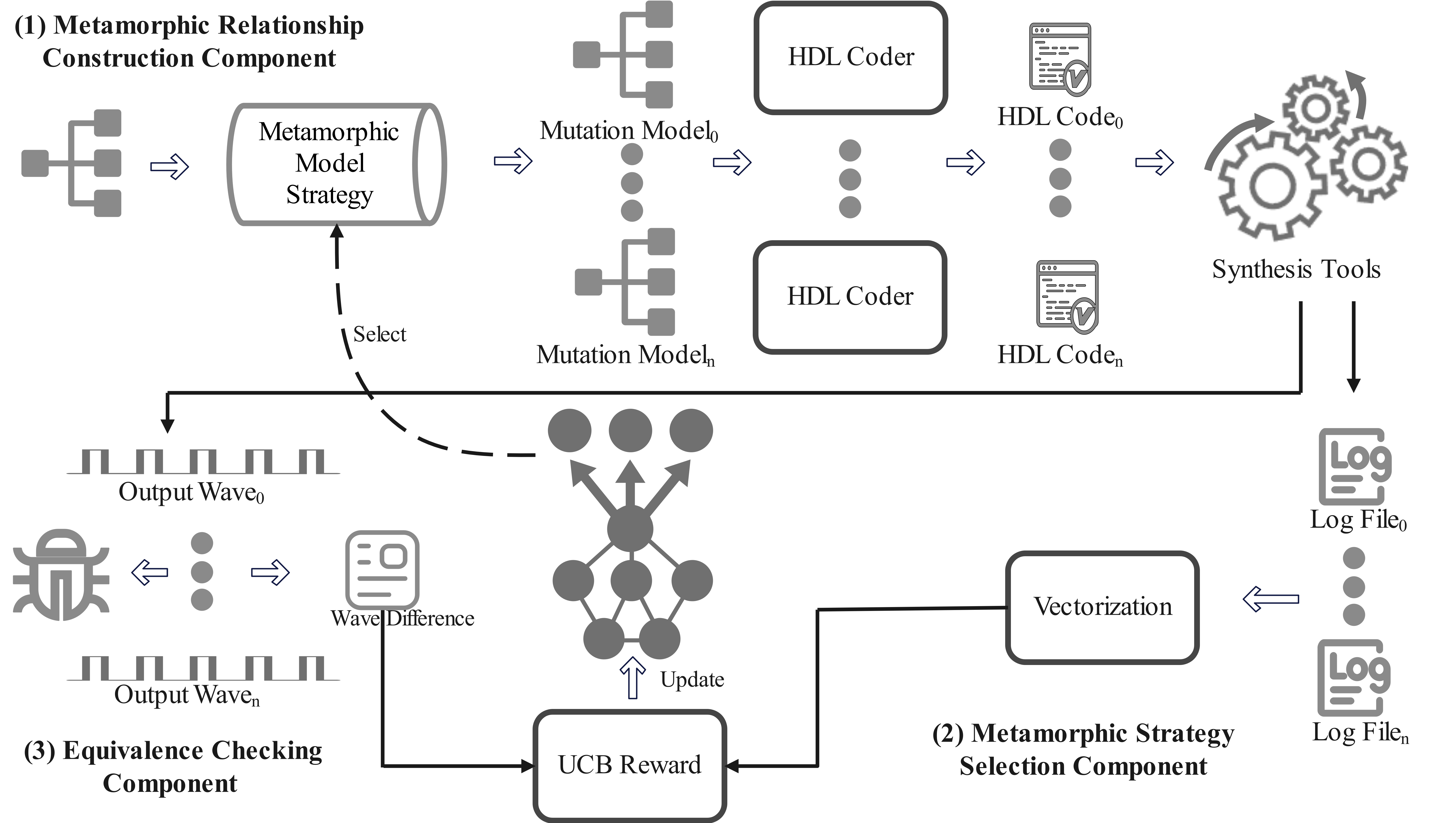}
\caption{Overview of Lin-Hunter}
\label{fig:figure2}
\end{figure}

\section{Methodology}
\label{sec:approach}

In this section, we first present an overview of the framework of Lin-Hunter. We then explain the Metamorphic relationship construction component and the Metamorphic strategy selection component to address the challenges of efficiency and diversity. Finally, the details of differential testing component are presented.

\subsection{Overview}
\label{sec:overview}

The framework of Lin-Hunter has been illustrated in Figure \ref{fig:figure2}.
Lin-Hunter includes three components: the Metamorphic relationship construction component, the Metamorphic strategy selection component, and the differential testing component.
The basic idea of Lin-Hunter is to use metamorphic relationships to construct diversity HDL test cases and use the Lin\_UCB algorithm to guide the metamorphic relationship construction strategies selection to enhance bug discovery efficiency.
Specifically, the equivalent mutation component designs four metamorphic relationship construction strategies (e.g., inserts assertion statements) to generate output equivalent but structurally different test case to enhance the diversity of HDL test cases.
The Metamorphic strategy selection component employs the Lin\_UCB algorithm to guide the selection of metamorphic construction strategies by analyzing the synthesis log information of the test case. 
To find more bugs in shorter time, the metamorphic strategy selection component aims to set higher reward of Metamorphic strategy  that are more likely to trigger bugs.
Finally, the differential testing component compares the output consistency between the HDL test case which has metamorphic relationships to detect bugs in FPGA logic synthesis compiler. 
If the output of the HDL test case which has metamorphic relationships is inconsistent, it indicates a bug in the FPGA logic synthesis compiler.

\subsection{Metamorphic Relationship Construction Component}

The metamorphic relationship construction component begin with the seed CPS models, which are generated by automatically Simulink model generation tools and real-word CPS corpus. 
Then the Metamorphic relationship construction component will generate diversity HDL test cases by applying metamorphic relationship construction rules to the seed CPS models. 
The metamorphic relationship construction strategies like inserting CPS block in non-executing regions, CPS data breakpoint recovery, will not change the final output of original seed CPS models but enhance the diversity of CPS models and finally enhance the diversity of HDL test cases. 
The new generated CPS models which has metamorphic relationships with original seed CPS models are called mutant models. 
Then the Metamorphic relationship construction component will generate HDL test cases by converting the mutant models into HDL code using HDL Coder. The HDL code will be used to test the FPGA logic synthesis compiler. 
The Metamorphic relationship construction component will also record the log information of the HDL code synthesis process. 
The log information will be used by the Metamorphic strategy selection component to guide the selection of mutation strategies.
Specifically, the metamorphic relationship construction strategies has been lines as follows.

\subsubsection{\textbf{Inserting CPS blocks in non-executing regions.}}
The first strategy to construct metamorphic relationships is to insert CPS modules in non-executing regions. 
The process can be formalized as formula~\ref{eq:insert}.
We define the original program as \( P \) with input \( x \) and output \( y = P(x) \). Let \( R_{non} \) denote a non-executing region of the program, satisfying the conditions as formula~\ref{eq:non}.
\begin{equation}
  \label{eq:non}
\forall x, \quad R_{non}(x) = \text{False}
\end{equation}

It means there is no execution path under any input \( x \) will trigger the code within \( R_{non} \).
By inserting CPS blocks \( b \) into \( R_{non} \), we obtain a metamorphic variant \( P' \).

\begin{equation}
  \label{eq:insert}
P'(x) = P(x) + b, \quad \text{where} \quad b \subset R_{non}
\end{equation}

The output is equivalent between \( P' \) and \( P \)  because the inserted CPS module \( C \) resides entirely within a non-executing region, ensuring that for any input \( x \), the execution path of \( P' \) remains identical to that of \( P \). 
Consequently, the observable outputs of the program are unaffected by the insertion, thereby preserving the program's functional behavior while enabling controlled metamorphic transformations.

\subsubsection{\textbf{Inserting if-else statements at random points.}}
The second strategy to construct metamorphic relationships is to insert \texttt{if-else} statements at arbitrary locations in the program. The process can be formalized as formula~\ref{eq:ifelse}.

\begin{equation}
  \label{eq:ifelse}
  P'(x) = P(x) + S,
  \end{equation}
where \( S \) denotes the inserted \texttt{if-else} blocks and inserted blocks, which can be formalized as formula~\ref{eq:ifcondition}.

\begin{equation}
  \label{eq:ifcondition}
  S = \texttt{if}~\texttt{\{}~C_1~\texttt{\}}~\texttt{else}~\texttt{\{}~C_2~\texttt{\}}.
  \end{equation} 
where \( C_1\) are the subsequent blocks of original seed models and \( C_2 \) are the randomly generated blocks. 
The condition of this assertion statement will be set as true, so the output of the program will not be changed. 
Furthermore, this ensures that neither \( C_1 \) nor \( C_2 \) produces side effects that influence \( P \)'s control flow, data flow, or outputs. 
As a result, the observable behavior of the program is preserved, allowing safe and controlled application of metamorphic transformations.

\subsubsection{\textbf{Promoting certain regions to subsystems.}}
The third strategy to construct metamorphic relationships is to promote certain regions of the program to independent subsystems. The process can be formalized as formula~\ref{eq:subsystem}.

\begin{equation}
  \label{eq:subsystem}
  P'(x) = P(x) \setminus R + \texttt{Sub}(R),
\end{equation}
where \( R \) denotes the selected region of the original program, and \( \texttt{Sub}(R) \) represents an encapsulated subsystem constructed from \( R \). 
Specifically, the subsystem \( \texttt{Sub}(R) \) is invoked in place of \( R \), while preserving the same input-output behavior as the original region, which can be formalized as formula~\ref{eq:subequiv}.

\begin{equation}
  \label{eq:subequiv}
  \forall x, \quad R(x) = \texttt{Sub}(R)(x).
\end{equation}

The execution of the original CPS model \(p\) has not changed because the subsystem \( \texttt{Sub}(R) \) is a functional equivalent of the original region \( R \), and the replacement does not introduce side effects that modify the control flow, data flow, or outputs of \( P \).  
Furthermore, HDL coder will convert each subsystem as a dependent HDL file which has reference with the main HDL files. 
Hence this metamorphic relationship construction strategy can enhance the diversity of HDL test cases by enhancing the reference relationships.

\subsubsection{\textbf{Transferring certain regions to new models.}}
The fourth strategy to construct metamorphic relationships is to transfer certain regions of the program into newly created models. 
The process can be formalized as formula~\ref{eq:transfer}.

\begin{equation}
  \label{eq:transfer}
  P'(x) = P(x) \setminus R + M_R(x),
\end{equation}
where  \( R \) denotes a selected region of the original program, and \( M_R \) is a newly constructed model that replicates the functionality of \( R \). 
Specifically, \( M_R \) takes the same inputs as \( R \) and is designed to produce the same outputs, which can be formalized as formula~\ref{eq:modelequiv}.

\begin{equation}
  \label{eq:modelequiv}
  \forall x, \quad R(x) = M_R(x).
\end{equation}

This equivalence guarantees that replacing \( R \) with \( M_R \) does not affect the correctness of the computation within the program.

Furthermore, to ensure that the overall output of the program remains unchanged, the model \( M_R \) will satisfy several constraints:
\begin{itemize}
  \item \( M_R \) must preserve the data dependencies of \( R \), ensuring that all inputs and outputs are consistent with the original program's execution.
  \item \( M_R \) must execute without introducing additional side effects that could alter the global program state, such as modifying shared variables or altering control flow beyond the boundaries of \( R \).
  \item The integration of \( M_R \) must maintain synchronization with the remaining components of \( P \), so that the interactions between \( M_R \) and the other parts of the program are identical to those between \( R \) and the rest of the program in the original implementation.
\end{itemize}

As a result, under these constraints, the replacement of \( R \) with \( M_R \) preserves the observable behavior of the program, ensuring that the final outputs of \( P' \) remain identical to those of \( P \) for any input \( x \).

To enhance the bug-finding capability of individual strategies, which only make limited modifications to the original CPS seed model, Lin-Hunter combines these strategies during execution. 
The combination of strategies aims to improve the diversity of the generated HDL. 
Specifically, each strategy is assigned an initial weight, representing the probability of Lin-Hunter selecting that strategy. 
Different models exhibit various CPS characteristics, which are reflected in the HDL test cases, such as having more mathematical modules or fewer conditional jumps. 
Therefore, strategy selection needs to be adjusted for different models to cover more boundary conditions in testing. We choose to use reinforcement learning to guide the selection of metamorphic relationship strategies which has been illustrated in section \ref{section:Metamorphic strategy selection component}.

\subsection{Metamorphic Strategy Selection Component}
\label{section:Metamorphic strategy selection component}    

Prior to introducing our approach, we first analyze why Lin\_UCB is suitable for FPGA synthesis testing. 
Multi-agent methods (such as DQN, PPO) can fit complex nonlinear relationships but have high computational complexity and slow convergence~\cite{hu2024review}. 
Many features of FPGA synthesis (such as the number of LUTs, path delay) have a significant impact on final performance and often exhibit linear or approximately linear relationships. 
The linear assumption of LinUCB is reasonable in this scenario and avoids overfitting. 
Additionally, LinUCB is lightweight and does not require complex multi-agent collaborative control, which reduces computational complexity.
Theoretically, complex methods not only fail to improve testing efficiency but also make the testing process overly complicated. 
Lightweight methods can balance testing efficiency and computational complexity. 
Therefore we choose LinUCB as the core algorithm for metamorphic relationship strategies selection.

Specifically, the Metamorphic strategy selection component, as illustrated in algorithm~\ref{alg:modified-linucb}, treats each metamorphic relationship strategy as an arm of a multi-armed bandit. 
And the Metamorphic strategy selection component will maintain and update a reward estimation model that considers both the immediate rewards from bug detection and a penalty term for repeatedly found bugs, allowing us to balance between exploiting successful strategies and exploring potentially valuable new ones. 
By consistently exploring new strategies and exploiting successful ones, the Metamorphic strategy selection component can guide the selection of metamorphic relationship strategies untill reached the max round we set.

\begin{algorithm}[!t]
  \caption{Lin\_UCB for Equivalence Mutation Strategy Selection}
  \label{alg:modified-linucb}
  \begin{algorithmic}[1]
  \Statex \textbf{Input:} Exploration parameter $\alpha$, penalization factor $\beta$, total rounds $T$, context vectors $\mathbf{x}_a$, penalty factors $f_a$ for all strategies $a$. 
  \Statex \textbf{Output:} Selected strategies $\{a_t\}_{t=1}^T$
  
  \State Initialize $\mathbf{A}_a \gets \mathbf{I}_d$, $\mathbf{b}_a \gets \mathbf{0}_d$ for all strategies $a$
  \State Set exploration parameter $\alpha > 0$ and penalization factor $\beta > 0$
  \For{each round $t = 1, 2, \dots, T$}
      \State Observe context vector $\mathbf{x}_a$ for each strategy $a$
      \For{each strategy $a$}
          \State Compute $\hat{\mathbf{\theta}}_a \gets \mathbf{A}_a^{-1} \mathbf{b}_a$
          \State Compute reward estimate $\hat{r}_a \gets \mathbf{x}_a^\top \hat{\mathbf{\theta}}_a$
          \State Adjust reward: $\hat{r}_a' \gets \hat{r}_a \cdot \exp(-\beta f_a)$
          \State Compute UCB: $\text{UCB}_a \gets \hat{r}_a' + \alpha \cdot \sqrt{\mathbf{x}_a^\top \mathbf{A}_a^{-1} \mathbf{x}_a}$
      \EndFor
      \State Select strategy $a_t \gets \arg\max_a \text{UCB}_a$
      \State Apply strategy $a_t$ and observe reward $r_t$
      \State Update $\mathbf{A}_{a_t} \gets \mathbf{A}_{a_t} + \mathbf{x}_{a_t} \mathbf{x}_{a_t}^\top$
      \State Update $\mathbf{b}_{a_t} \gets \mathbf{b}_{a_t} + r_t \mathbf{x}_{a_t}$
  \EndFor
  \State Return $\{a_t\}_{t=1}^T$
  \end{algorithmic}
  \end{algorithm}

The selection process begin with two initialization steps (lines 1-2 in Algorithm \ref{alg:modified-linucb}). 
Firstly, the covariance matrix \(\mathbf{A}_a\) is initialized as the identity matrix \(\mathbf{I}_d\), ensuring initial feature independence. The cumulative reward vector \(\mathbf{b}_a\) is initialized as zero vector \(\mathbf{0}_d\). 
Secondly, the exploration parameter \(\alpha\) is set to 1.0 based on empirical studies showing optimal trade-off in similar contextual bandit problems. 
The penalization factor \(\beta\) is set to 0.5, determined through ablation studies comparing \(\beta \in \{0.1, 0.3, 0.5, 0.7, 0.9\}\).

Then, each metamorphic relationship strategy $a$ will be represented by a feature vector $\mathbf{x}_a$, which encodes contextual information. 
The contextual information includes the strategy type, the historical performance metric, and the occurrence frequency $f_a$ of previously discovered bugs. The historical performance metric is the number of bugs detected by the metamorphic relationship strategy normalized to \([0, 1]\). 
The occurrence frequency $f_a$ is calculated by dividing the repetition counter $C_i$ by the total rounds $T$. 
For example, for strategy \texttt{insert CPS blocks in non-executing regions}, the metamorphic relationship strategy type will be $[1,0,0,0]$ and the historical performance metric is $0.7$, and the occurrence frequency $f_a$ of previously discovered bugs is $0.2$. 
Hence the $\mathbf{x}_a$ will be $[1,0,0,0,0.7,0.2]$.

\begin{equation}
  \hat{\mathbf{\theta}}_a = \mathbf{A}_a^{-1}  \cdot \mathbf{b}_a
  \label{eq:1}
  \end{equation}
  
Due to the linear relationship between the features and the reward, the metamorphic strategy selection component will compute the linear model parameters \(\mathbf{\theta}\) of strategy $a$ to estimate the reward distribution of the current strategy. 
Specifically, \(\mathbf{\theta}\) is a product of covariance matrix \(\mathbf{A}_a^{-1}\) and cumulative reward vector \( \mathbf{b}_a\).
Hence the linear model parameters \(\mathbf{\theta}\) of strategy $a$ can be calculated as shown in Equation~\ref{eq:1}.

  \begin{equation}
    \label{eq:2}
    \hat{r}_a = \mathbf{x}_a^\top \cdot \hat{\mathbf{\theta}}_a
    \end{equation}
    
Then, the metamorphic strategy selection component use linear model parameters to estimate the expected reward under the current policy $a$, which is the dot product of the current context features $\mathbf{x}_a$ and the model parameters \(\mathbf{\theta}\) as shown as formula~\ref{eq:2}.
This approach allows for a rapid estimation of the potential reward for each strategy combination, facilitating the selection of the optimal one for execution

\begin{equation}
  \label{eq:3}
\hat{r}_a' = \hat{r}_a \cdot \exp(-\beta f_a)
\end{equation}

The metamorphic strategy selection component aims to adjust the reward $\hat{r}_a$ to discovery of new bugs.
Hence, the metamorphic strategy selection component chose to include a penalty term $\beta$ in the reward adjustment process.
So, the adjusted reward $\hat{r}_a'$ is calculated as the product of the estimated reward $\hat{r}_a$ and the exponential of the negative product of the penalization factor $\beta$ and the bug occurrence frequency $f_a$ as shown as formula \ref{eq:3}.

Moreover, the metamorphic strategy selection component defines FPGA synthesis bugs in two categories: inconsistency bugs and crash bugs. 
Inconsistency bugs occur when the output from different logic synthesis tools is inconsistent, while crash bugs occur when the logic synthesis tools unexpectedly crash during the synthesis of HDL test cases. 
Inconsistency bugs require manual review, although there are some tools that can automatically reduce the inconsistency HDL code.
But crash bugs will generate a log that provides information about the root cause of the bug. Therefore, for automatically bugs detection, we choose to analyze the log information of crash bugs.
That is why our method can find more unknown crash bugs in logic synthesis tools.
% Hence based the analysis of the bug log info, the metamorphic strategy selection component set a dynamic reward update algorithm to motivate the metamorphic strategy selection component find more new bugs not only find repeated bugs.
% To find more bugs, the metamorphic strategy selection component employs a dynamic reward update algorithm based on the analysis of bug log information, motivating it to discover new bugs rather than just repeated ones.
According to the suggestions of the previous work on log analysis, the metamorphic strategy selection component trained a Word2vec~\cite{word2vec} model to extract specific information from the bug log, such as the bug memory address, the name of the function that triggered the bug, and vectorized them. 
Then, the metamorphic strategy selection component used cosine similarity to calculate the similarity between bugs. The cosine similarity formula is shown in Equation~\ref{eq:cos}, where $\mathbf{u}$ and $\mathbf{v}$ are two vectors representing the features of two bugs.
\begin{equation}
     \cos(\theta) = \frac{\mathbf{u} \cdot \mathbf{v}}{\|\mathbf{u}\| \|\mathbf{v}\|}
     \label{eq:cos}
\end{equation}

Then the metamorphic strategy selection component uses K-Means~\cite{kmeans} to cluster found bugs and existing bugs, which use cosine similarity as the distance metric. 
This can determine whether the bug falls into a known bug class. 
In addition, the metamorphic strategy selection component will maintain a set repetition counter $C_i$ to record the number of repetitions of a bug class. 
$f_a$ is the frequency of bugs which calculated by  repetition counter $C_i$ divide total rounds $T$

% \textbf{UCB Computation and Strategy Selection (lines 9-12 in Algorithm \ref{alg:modified-linucb})}

\begin{equation}
  \label{eq:4}
\text{UCB}_a = \hat{r}_a' + \alpha \cdot \sqrt{\mathbf{x}_a^\top \mathbf{A}_a^{-1} \mathbf{x}_a}
\end{equation}

After getting the adjusted reward $\hat{r}_a'$ (lines 9-12 in Algorithm \ref{alg:modified-linucb}), the metamorphic strategy selection component will compute the UCB score which will be used as metric of strategy selection.
The UCB score for each strategy $a$ computed as the sum of the reward estimate and the width of the confidence interval which has shown in formula~\ref{eq:4}. 
The width of the confidence interval will be presented as $\alpha \cdot \sqrt{\mathbf{x}_a^\top \mathbf{A}_a^{-1} \mathbf{x}_a}$, which will be used to encourages exploration of strategies with higher uncertainty.
At each round, the strategy with the highest UCB score is selected. 
Then, the metamorphic strategy selection component applies strategy $a_t$ and gets reward $r_t$.

% % This shows that the metamorphic strategy selection component determines the best strategy for the current round based on its potential reward and confidence interval.
% \textbf{Dynamic Updates (lines 13-14 in Algorithm \ref{alg:modified-linucb})}
\begin{equation}
  \label{eq:5}
  \mathbf{A}_{a_t} \gets \mathbf{A}_{a_t} + \mathbf{x}_{a_t} \mathbf{x}_{a_t}^\top, \quad
  \mathbf{b}_{a_t} \gets \mathbf{b}_{a_t} + r_t \mathbf{x}_{a_t}
  \end{equation}

Finally, as shown as formula ~\ref{eq:5}, to adjust the model parameters with new reward data so that it gradually reflects the true reward distribution, the metamorphic strategy selection component updates the covariance matrix $\mathbf{A}_{a_t}$ of the strategy $a_t$ (lines 13-14 in Algorithm \ref{alg:modified-linucb}).

% It aims to record new context feature information and gradually adjust the correlation between features.
% Meanwhile, Lin-Hunter also updates the cumulative reward vector \(\mathbf{b}_{a_t}\) of the strategy $a_t$. 
% This can be used to adjust the model parameters with new reward data so that it gradually reflects the true reward distribution.

% % The metamorphic strategy selection component updates the covariance matrix \(\mathbf{A}_{a_t}\) of the strategy \(a_t\) and the cumulative reward vector \(\mathbf{b}_{a_t}\) of the strategy \(a_t\).

After updating all parameters, Lin-Hunter will perform the next round of selection.
The method continues to iterate until it reaches the round $T$ Lin-Hunter set. 
By doing so, Lin-Hunter chooses better strategies in continuous iterations, which makes Lin-Hunter to find more bugs in a shorter time period and discover bugs that we have not discovered before. 
Due to the different quality of test case generation, the optimal equivalence mutation strategy is different for different test cases, so we do not recommend the best configuration. 
However, according to our multiple experiments, the strategy of \texttt{transferring certain regions to new models} is the most balanced strategy. 
This is also because increasing the complexity of the reference relationship of the test cases can better touch the test boundary of logic synthesis.

\begin{figure}[t]
\centering  
%\includesvg[width=\linewidth]{Difftest2.svg}
\includegraphics[width=0.9\linewidth]{ 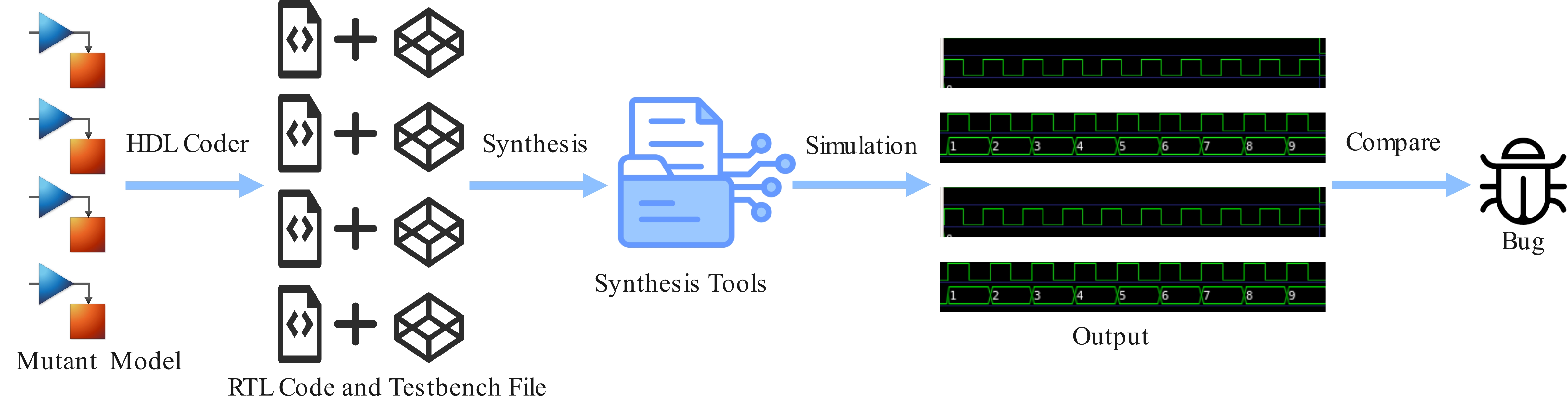}
\caption{Differential Test Process}
\label{fig:figure4}
\end{figure}

\subsection{Differential Testing Component}

The differential testing component is used to detect bugs within FPGA logic synthesis tools. 
The basic idea of the differential testing is to compare the output netlists of the test cases which has metamorphic relationships to identify potential bugs.
Specifically, the differential testing component will input the HDL test case which has metamorphic relationships into different FPGA logic synthesis tools and perform synthesis. 
Then, the differential testing component will compare the output netlists of the test cases which has metamorphic relationships.
If the output netlists of the test cases which has metamorphic relationships are inconsistent, it indicates a bug in the FPGA logic synthesis compiler.
The differential testing component will record the bug information. This process is depicted in figure~\ref{fig:figure4}.
We use the SymbiYosys of Yosys, which leveraging SAT and SMT solvers to confirm netlists consistency. 
To explicitly identify the root cause of the bug, Lin-Hunter simplified the bug-triggering HDL test case variants using both automated and manual reduction approaches.
Specifically, Lin-Hunter performed an iterative removal of modules, assignments, and logic blocks from the model.
If the bugs persisted after removing a module, that module was deemed unrelated to the faults. 
Otherwise, the block was restored to its original position. 
This process was repeated until no additional blocks could be removed.
To avoid reporting duplicate bugs, Lin-Hunter manually leveraged failed assertions and backtracking to identify duplicates.
When two bugs have the same failed assertion or backtracking, Lin-Hunter consider them duplicates.
Finally, Lin-Hunter submitted the detected bugs as \texttt{new} bugs to FPGA logic synthesis tools Support website.

\section{Evaluation}
\label{sec:Evaluation}

\subsection{Experimental Setup}
Lin-Hunter is used by Matlab and Python, running on a server with Ubuntu 22.04, equipped with an Intel Core i9 CPU @2.10GHz and 128GB of memory. 
To evaluate the effectiveness of Lin-Hunter, we utilized four testing tools: \textbf{Vivado}, \textbf{Yosys}, \textbf{Iverilog}, and \textbf{Quartus}. 
We had tested four synthesis tools, including the open-source tools Yosys and Iverilog, as well as commercial software Vivado and Quartus. 
Yosys serves as a synthesis tool, Iverilog as a simulation tool, while Vivado and Quartus integrate both synthesis and simulation functionalities. By using the latest versions of these tools, we validated the effectiveness and reliability of the fuzzer.
For the Lin\_UCB algorithm, we set the exploration parameter \(\alpha\) to 1.0 to balance exploration and exploitation, and the penalization factor \(\beta\) to 0.5 to moderately penalize repeated bugs while maintaining algorithm stability.

\subsection{Baselines}
We chose LegoHDL~\cite{LegoHDL} as our baseline, which is the state-of-the-art FPGA logic synthesis compiler fuzzer. LegoHDL discovered 20 bugs within three months, demonstrating its bug-finding capabilities. 
We reproduced it according to the instructions in its GitHub README and used its default configuration. 
Additionally, we selected Verismith~\cite{Verismith} as another baseline because it is one of the most popular fuzzing test methods for FPGA logic synthesis tools, having found 11 bugs over two years, showing its effectiveness. 
We reproduced it according to the instructions in its GitHub README and used its default configuration.

\subsection{Seed Test Cases}
Our dataset includes two parts of CPS models to better assist us in generating complex and diverse HDL test cases. 
The first part is generated by the automated tool LegoHDL~\cite{LegoHDL}, and the second part is collected from real-world models on open-source platforms like GitHub\footnote{\url{https://github.com/verivital/slsf_randgen/wiki/Curated-Corpus-of-Publicly-Available-Simulink-Models}}. 
LegoHDL-generated models offer high complexity and customization, while open-source models provide diversity from practical applications.

\subsection{Research Questions}

In this section, we investigate four research questions (RQs) to evaluate the effectiveness of Lin-Hunter. Specifically, our evaluation aims to answer the following three research questions:

\begin{itemize}[leftmargin=*]
    \item \textbf{RQ1}: How effective is Lin-Hunter in identifying bugs in FPGA logic synthesis tools?
    \item \textbf{RQ2}: How does the performance of Lin-Hunter compare to the current state-of-the-art FPGA fuzzers?
    \item \textbf{RQ3}: Is the use of reinforcement learning algorithms beneficial for selecting CPS mutation strategies?
    \item \textbf{RQ4}: Is the Lin\_UCB algorithm in Lin-Hunter effective in finding bugs in FPGA logic synthesis tools?
\end{itemize}

% \textbf{RQ1}: How effective is Lin-Hunter in identifying bugs in FPGA logic synthesis tools?
% \textbf{RQ2}: How does the performance of Lin-Hunter compare to the current state-of-the-art FPGA fuzzers?
% \textbf{RQ3}: Is the use of reinforcement learning algorithms beneficial for selecting CPS mutation strategies?
% \textbf{RQ4}: Is the Lin\_UCB algorithm in Lin-Hunter effective in finding bugs in FPGA logic synthesis tools?

% RQ1 is used to assess the advantage of Fuzz-TK in statement-level fault localization. RQ2 evaluates the contribution of each component in Fuzz-TK's performance. RQ3 aims to evaluate the impact of different fuzzy sets on Fuzz-TK. RQ4 aims to evaluate the impact of different training processes on Fuzz-TK. RQ5 aims to evaluate the impact of different training set sizes on Fuzz-TK.
RQ1 is used to evaluate the effectiveness of Lin-Hunter in testing FPGA logic synthesis tools. RQ2 assesses the performance of Lin-Hunter in comparison to state-of-the-art (LegoHDL) and popular (Verismith) FPGA logic synthesis tools fuzzing test methods.RQ3 aims to assess how effectively the Lin\_UCB algorithm, based on reinforcement learning, can guide the mutation strategies for CPS models. RQ4 aims to evaluate the impact of Lin-Hunter's most critical component, Lin\_UCB, on the overall methodology.

\subsection{Answer to RQ1}

We conducted experiments from August 2024 to November 2024 to evaluate Lin-Hunter's bug-finding capabilities, using the latest versions of Vivado\footnote{\url{https://adaptivesupport.amd.com/s/topic/0TO2E000000YKY5WAO/synthesis?language=en_US}} 2024.1, Yosys\footnote{\url{https://github.com/YosysHQ/yosys/issues}} 0.46, Icarus Verilog\footnote{\url{https://github.com/steveicarus/iverilog/issues}} 12.0, and Quartus\footnote{\url{https://community.intel.com/}} 24.1. 
Through bisection reduction, we manually reduced the test cases that triggered the detected bugs to the smallest reproducible Verilog bug . We use a binary search method to iteratively simplify the code. 
By progressively removing code blocks, commenting out signals, and reducing functional logic, we aim to identify the minimal code segment that triggers the bug.
Simulation validation is conducted using the same testbench across FPGA logic synthesis tools to verify whether the bug persists. Finally, all detected bugs were submitted to the relevant communities.

\begin{table}[t]
\centering
\caption{The details of bugs found by Lin-Hunter}
\label{table:bugs-1}
\begin{tabularx}{\linewidth}{c l X c c c}
\toprule
\textbf{Num} & \textbf{ID} & \textbf{Summary} & \textbf{Status} & \textbf{Type} & \textbf{Software} \\ 
\midrule
1 & o5n7eSAA & HARTNlUtil::isCarryInst error & Verified & Confirmed & Vivado \\ 
2 & HZdyHSAT & NNetC::singleDriver error & Verified & Confirmed & Vivado \\ 
3 & HZe1QSAT & HARTGLAddGen::regenerate error & Verified & Confirmed & Vivado \\ 
4 & HrhKDSAZ & NPinC::parentModule error & Verified & Confirmed & Vivado \\ 
5 & gYVExSAO & DFPin::disconnect error & Verified & Confirmed & Vivado \\ 
6 & hHjB1SAK & HARTTUpdateTNInstC::Cell error & Verified & Confirmed & Vivado \\ 
7 & hHjCKSA0 & HARTXmsgWriter::Print error & Verified & Confirmed & Vivado \\ 
8 & iirCtSAI & dot::openFile error & Verified & Confirmed & Vivado \\ 
9 & iinFsSAI & GXorGen::bestSoln error & Verified & Confirmed & Vivado \\ 
10 & pZyQeSAK & ConstProp::reconnect error & Verified & Confirmed & Vivado \\ 
11 & pZzlhSAC & PrioMuxInfo::setPinArray error & Verified & Confirmed & Vivado \\ 
12 & 4O8t1SAC & ConstProp::propagate error & Verified & Confirmed & Vivado \\ 
13 & 4O9NaSAK & DFNode::calcConstantBinaryInt error & Verified & Confirmed & Vivado \\ 
14 & 4O9NcSAK & HARTOptMux::createPartition error & Verified & Confirmed & Vivado \\ 
15 & 4O90OSAS & NDup::dupGlobalNames error & Verified & Confirmed & Vivado \\ 
16 & 4O9nOSAS & NTargetLibC::findCell error & Verified & Confirmed & Vivado \\ 
17 & 4O9rQSAS & NBaseModC::realModule error & Verified & Confirmed & Vivado \\ 
18 & \#4697 & Sign extension of zero-width & Verified & Confirmed & Yosys \\ 
\bottomrule
\end{tabularx}
\end{table}

\textbf{Experiment Results.}
As demonstrated in Table~\ref{table:bugs-1}, Lin-Hunter discovered 18 bugs within 3 months, all of which have been officially confirmed, with 8 will be fixed in the latest versions. 
Additionally, all bugs can be reproduced from our GitHub homepage~\cite{Lin-Hunter}. Due to space limitations, we showcase two typical bugs in this paper.

\subsubsection{Bug1 Yosys \#4697: Sign Extension of Zero-Width Signals Causing Synthesis Inconsistency}
\begin{figure}[!t]
    \centering
    \begin{subfigure}[b]{0.35\textwidth}
        \centering
        \includegraphics[width=\textwidth]{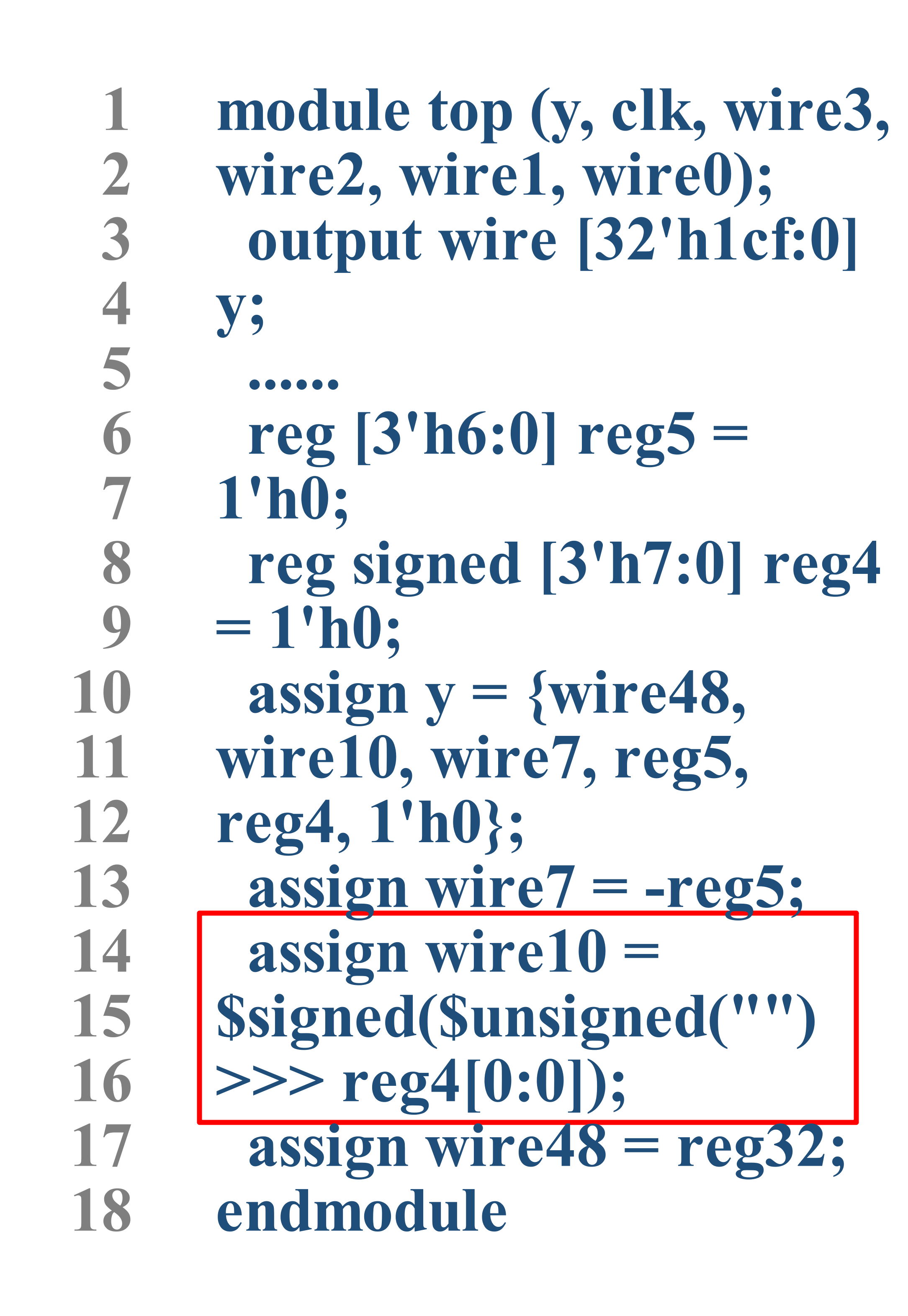}
        \caption{Bug \#4697 minimized Verilog code}
        \label{fig:sub-figure1}
    \end{subfigure}
    \hfill
    \begin{subfigure}[b]{0.65\textwidth}
        \centering
        \includegraphics[width=\textwidth]{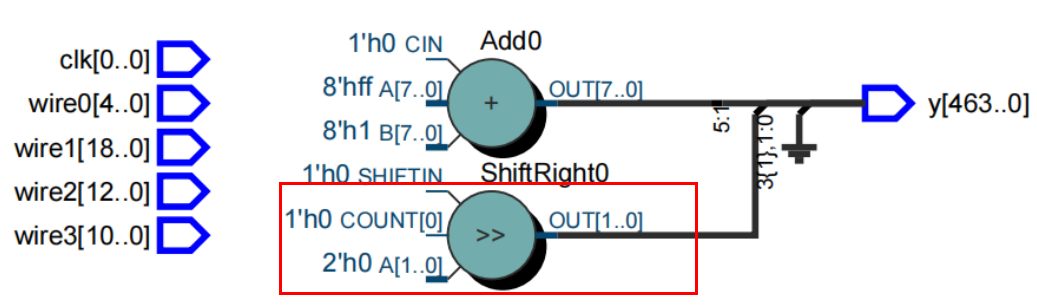}
        \caption{Netlist of Bug \#4697}
        \label{fig:sub-figure2}
    \end{subfigure}
    \caption{Reduced Example of Bug \#4697}
    \label{fig:figure5}
\end{figure}

We display a bug discovered by Lin-Hunter in the Yosys tool. 
Lin-Hunter simulated the original design and the netlist synthesized by Yosys and reported the differences. After our reduction process, the minimized Verilog code that triggers the bug is shown in the \figurename \ref{fig:sub-figure1}. This bug is caused by a shift operation.
 
Specifically, by analyzing, Lin-Hunter simulated the original design and the netlist synthesized by Yosys. 
The zero-width signal \{0\{1'b1\}\} was right-shifted and sign-extended, then assigned to a 4-bit output \textit{b}. 
Since the sign extension behavior of zero-width signals is undefined, FPGA logic synthesis tools may produce different results at different stages. 
During the Verilog code design phase, zero-width signals may be ignored or treated as specific values (like 0); 
while during the synthesis phase, the synthesis tool (such as Yosys) may treat zero-width signals as \textit{Sx} (unknown state). 
The final hardware behavior may exhibit unexpected behavior due to the inconsistency between the synthesized circuit and the simulation results, leading to system vulnerabilities or functional bugs.

\subsubsection{Bug2 Vivado 8HZdyHSAT: Synthesis Crash of Overly Complex Nested Ternary Operator}
\begin{figure}[t]
\centering  
%\includesvg[width=\linewidth]{Difftest2.svg}
\includegraphics[width=0.8\linewidth]{ 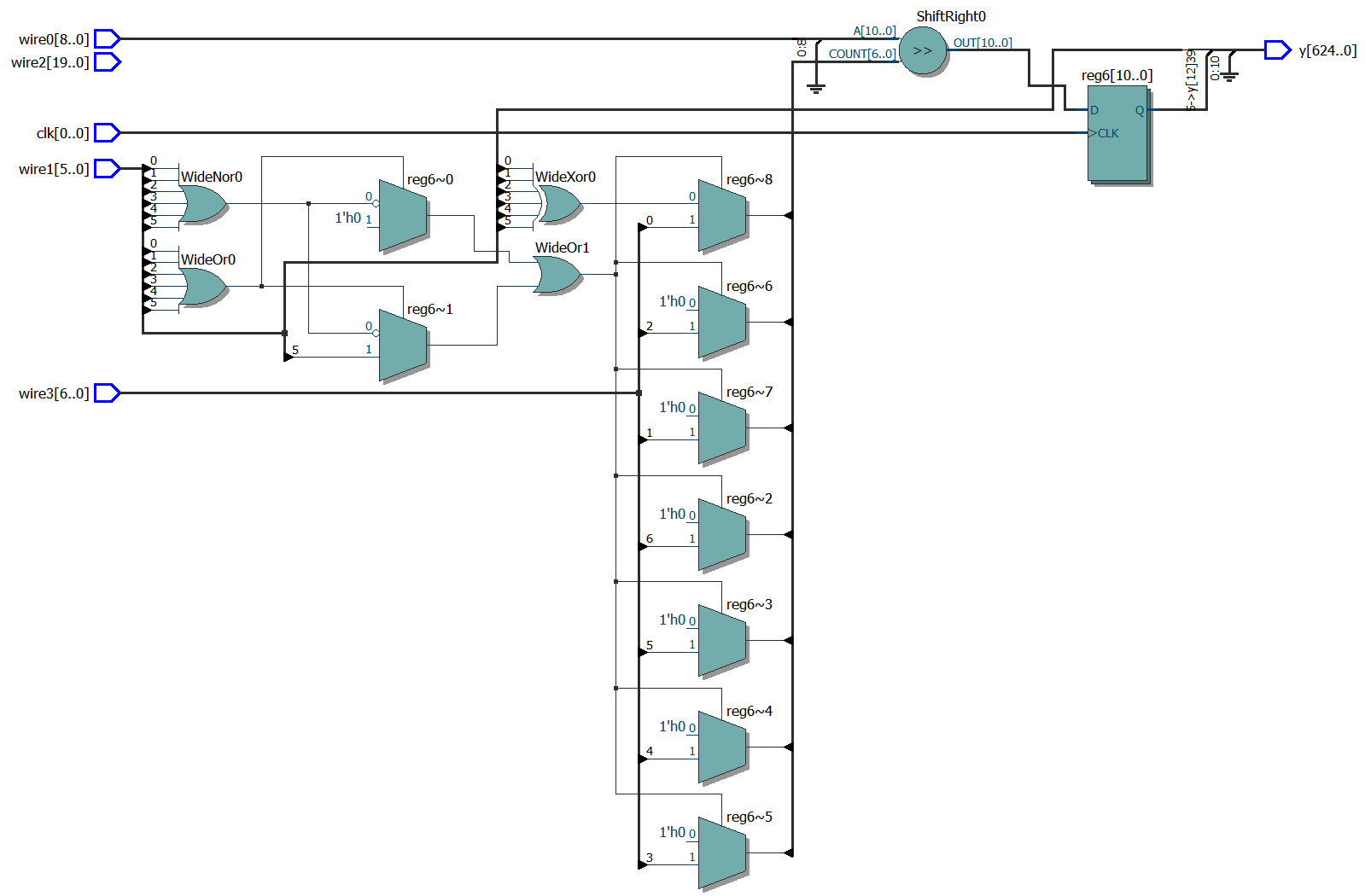}
\caption{Netlist of Bug 8HZdyHSAT}
\label{fig:figure8}
\end{figure}

As presented in \figurename \ref{fig:figure8}, the root cause of the Vivado crash is the overly complex nested ternary operator expression used in the always block of \textit{reg6}. This expression mixes signed and unsigned data types, bit selection, and shift operations, leading the Vivado synthesizer to crash as it struggles to effectively handle all possible scenarios during the parsing and optimization of this intricate combinational logic.

\textbf{Summary of RQ1.}
The experimental results indicate that Lin-Hunter is effective in identifying bugs in FPGA logic synthesis and simulation tools. Within 3 months, 18 valid bugs were discovered.

\subsection{Answer to RQ2}
Considering that developers typically aim to improve bug detection efficiency, we evaluated the effectiveness of Lin-Hunter by comparing its bug-finding capabilities with LegoHDL and Verismith. 
In this experiment, each method was used to generate the same number(2000 files) test cases with the same scale, based on the recommendation of Verismith we set the scale as 700-100 lines code. 
We then recorded the generation time of each method and compared the number of bugs which were found by each method within a week. By doing so, we compared the capability and efficiency of each method.

\textbf{Experiment Results.}
As can be seen in Table \ref{table:bugs}, during the fuzz testing of four tools (Vivado, Yosys, Iverilog, and Quartus), LegoHDL generated  \(2.7 \times 10^4\) test cases in 7 days, finding 2 bugs in Vivado, 1 bug in Yosys, 1 bug in Iverilog, and 1 bug in Quartus. Verismith generated  \(2.15 \times 10^4\) test cases in 7 days, finding 2 crash bugs in Vivado. Similar bug reports were found in the Vivado community, indicating these bugs were known.

In contrast, our proposed method, Lin-Hunter, generated  \(3.35\times 10^4\) test cases in 7 days, finding four new confirmed Vivado crash bugs, one confirmed Yosys inconsistency bug, one unconfirmed Yosys inconsistency bug, and one unconfirmed Quartus inconsistency bug.The unconfirmed bugs have been documented on our GitHub page~\cite{Lin-Hunter}. Lin-Hunter and LegoHDL outperformed the current popular method, Verismith, in bug detection. 

Lin-Hunter's superiority is attributed to its ability to access a broader corpus through the Simulink HDL block library. 
By generating ASTs to create CPS models and converting them into HDL code, Lin-Hunter employs a more comprehensive approach. 
The experiment also demonstrates Lin-Hunter's specific proficiency in detecting Vivado crashes, which can be attributed to the reward settings in the Lin\_UCB algorithm within Lin-Hunter. 
By analyzing error information in the logs of synthesized HDL code, Lin-Hunter effectively promotes the generation of HDL cases that trigger crashes.

\begin{table}[t]
\setlength{\tabcolsep}{4pt} % 仅对这个表格设置较小的列间距
\centering
\caption{Bugs Found by Lin-Hunter, LegoHDL, and Verismith}
\label{table:bugs}
\begin{tabular}{lccccc}
\toprule
\textbf{Approach} & \textbf{Vivado} & \textbf{Yosys} & \textbf{Iverilog} & \textbf{Quartus} & \textbf{Total} \\ 
\midrule
Verismith & 2 (Known) & 0 & 0 & 0 & 2 \\ 
LegoHDL & 2 & \textbf{1} & 1 & 1 & 5 \\ 
Lin-Hunter & \textbf{4} & \textbf{2} & \textbf{0} & \textbf{1} & \textbf{7} \\
\bottomrule
\end{tabular}
\end{table}

\setlength{\tabcolsep}{8pt} % 恢复全局设置

% \begin{figure}[t]
% \centering  
% %\includesvg[width=\linewidth]{Difftest2.svg}
% \includegraphics[width=0.95\linewidth]{ figs/fig7plus.png}
% \caption{Bugs Found by Different Approaches}
% \label{fig:bugs}
% \end{figure}

\begin{figure}[htbp]
    \centering
    \begin{subfigure}[b]{0.45\textwidth}
        \centering
        \includegraphics[width=\textwidth]{ 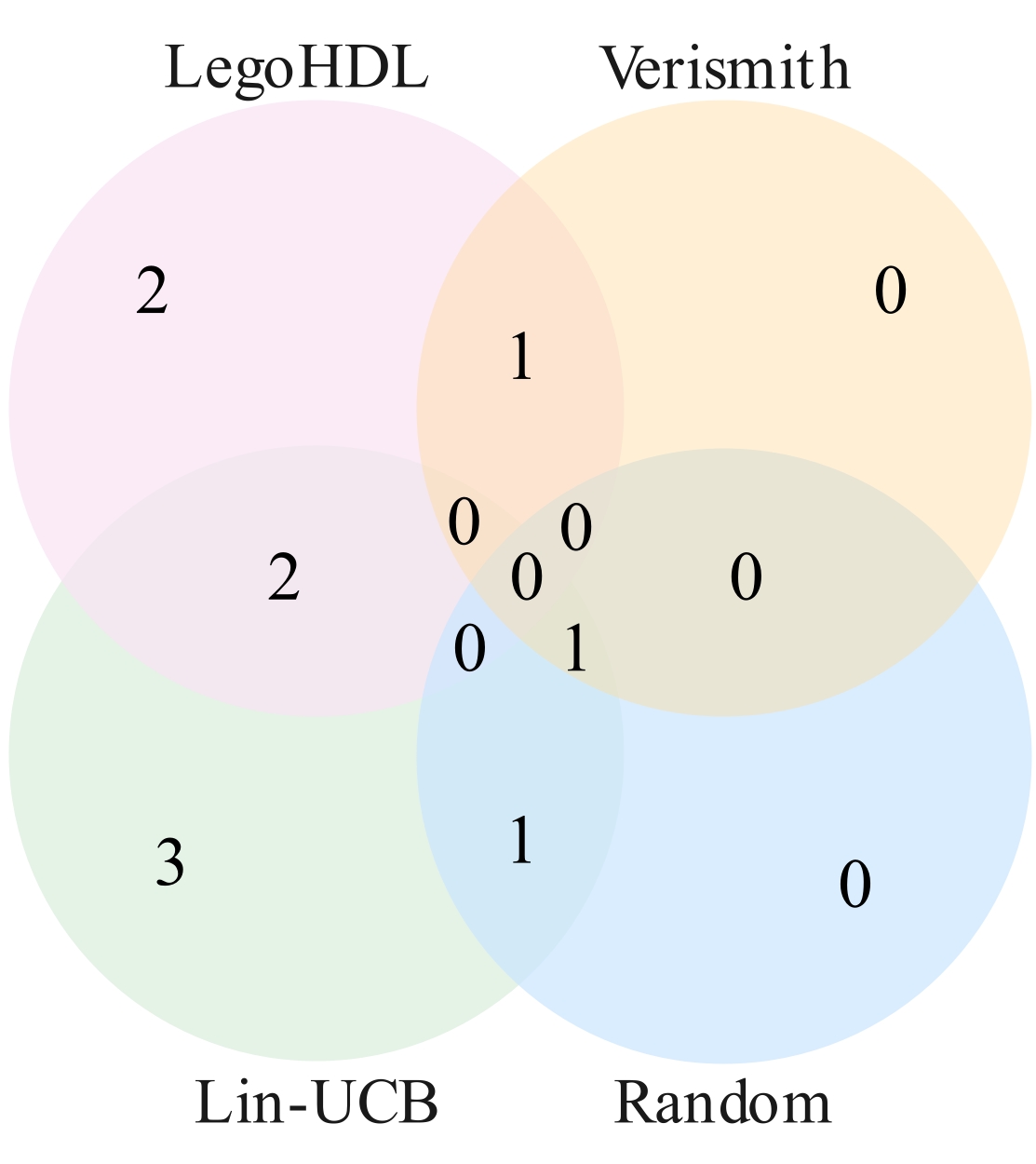}
        \caption{Number of Bugs}
        \label{fig:subfigC}
    \end{subfigure}
    \hfill
    \begin{subfigure}[b]{0.45\textwidth}
        \centering
        \includegraphics[width=\textwidth]{ 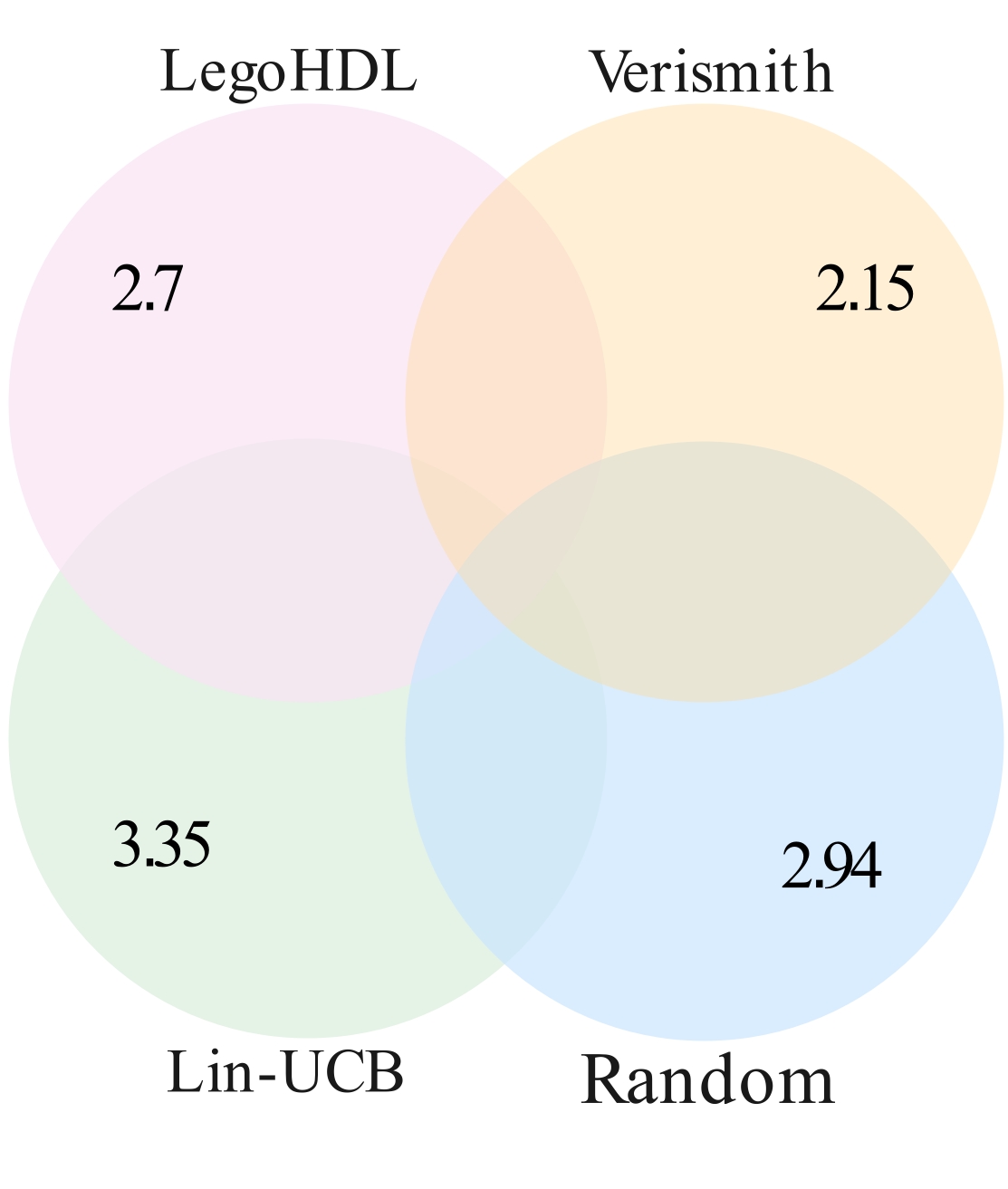}
        \caption{Number of HDL Cases ( $10^4$ )}
        \label{fig:subfigD}
    \end{subfigure}
    \caption{Comparison of HDL Cases and Bugs Found by Different Methods}
    \label{fig:figure6}
\end{figure}

\textbf{Summary of RQ2.}
As shown in Table \ref{table:bugs}, experimental results have demonstrated that Lin-Hunter has superior bug detection capabilities compared to Verismith. 
Additionally, the time required for Lin-Hunter to generate the same number of test cases was less than that of Verismith and LegoHDL. 
The test cases generated by Verismith appeared more redundant and meaningless, that is why it can not find more bugs compared to Lin-Hunter. 
Further more, Lin\_UCB algorithm has enhanced the efficiency of Lin-Hunter, that is why it can take less time compared to LegoHDL.

\subsection{Answer to RQ3}

Considering that Lin-Hunter, built on the foundation of LegoHDL, enhances the diversity of generated HDL test cases and further improves the testing efficiency of FPGA logic synthesis compilers, we conducted an ablation study to assess the effectiveness of Lin-Hunter's reinforcement learning-based Lin\_UCB algorithm in the final FPGA logic synthesis compiler testing. Specifically, we compared the efficiency of using Lin\_UCB and random strategies in selecting CPS model mutation strategies. Additionally, we reproduced the LegoHDL method (LegoHDL without model mutation). We tested Lin-Hunter continuously for 7 days under three methods. The evaluation metrics included the number of successfully generated HDL test cases and the number of detected vulnerabilities. For the number of successfully generated HDL test cases, we synthesized all generated HDL test cases and recorded the cases that could be correctly synthesized. For the number of detected vulnerabilities, we utilized an differential testing component to verify the generated HDL test cases and counted the vulnerabilities detected under each strategy.

\textbf{Experiment Results.} 
As illustrated in \figurename~\ref{fig:subfigD}, the use of Lin\_UCB achieved the best performance in terms of the number of generated HDL test cases and the ability to detect vulnerabilities. With the reinforcement learning algorithm Lin\_UCB, Lin-Hunter generated 13.95\% more HDL test cases than the random strategy and 24.07\% more than the LegoHDL method. Although Verismith generates Verilog code in milliseconds, the test cases it produces typically exceed 100KB in size. The high redundancy of these test cases results in significant time consumption during synthesis and simulation. Consequently, Verismith synthesizes the fewest test cases within a seven-day period. These experimental results highlight the advantage of the Metamorphic strategy selection component in enhancing the overall effectiveness of the framework. Guided by the Lin\_UCB algorithm, Lin-Hunter was able to generate more test cases that could be correctly synthesized, thereby demonstrating superior efficiency in vulnerability detection.

\textbf{Summary of RQ3.} 
The Metamorphic strategy selection component plays a critical role in improving the efficiency of HDL test case generation and subsequent vulnerability detection. By generating more complex and diverse test cases, this component significantly enhances Lin-Hunter's ability to identify challenging edge cases and ensure effective synthesis.

\begin{figure}[!t]
    \centering
    \begin{subfigure}[b]{0.4\textwidth}
        \centering
        \includegraphics[width=\textwidth]{ 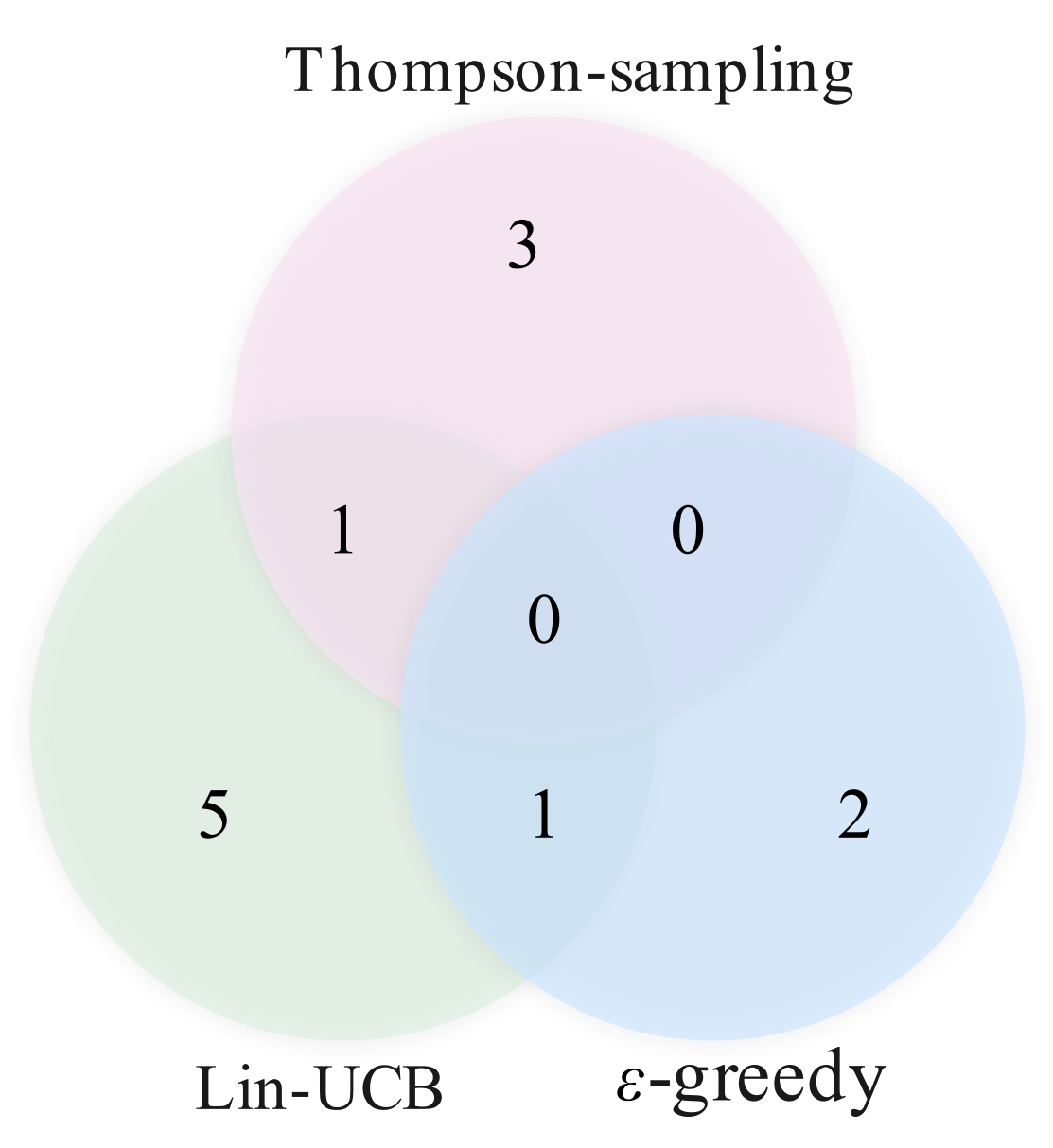}
        \caption{Number of Bugs}
        \label{fig:subfigA}
    \end{subfigure}
    \hfill
    \begin{subfigure}[b]{0.4\textwidth}
        \centering
        \includegraphics[width=\textwidth]{ 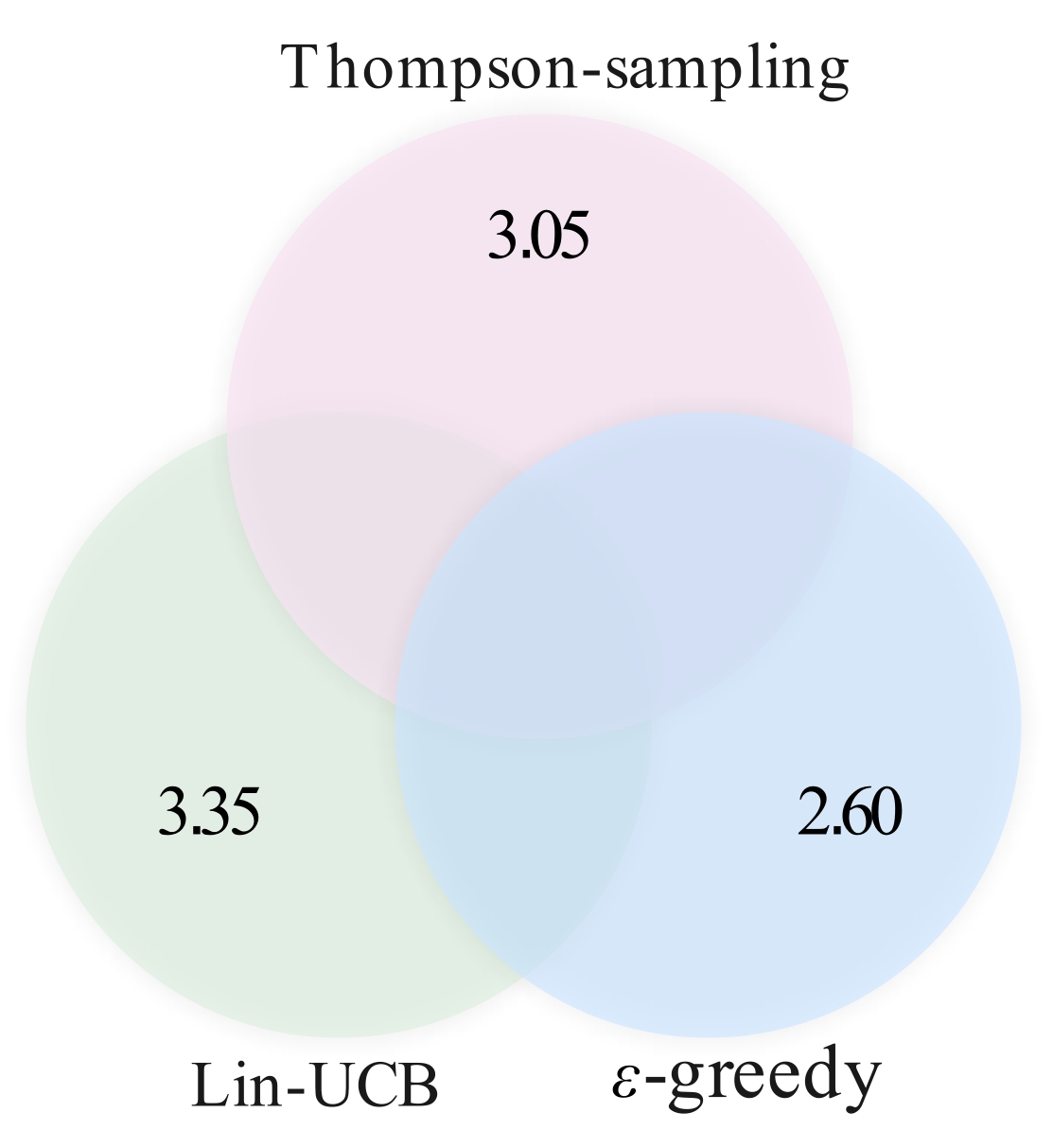}
        \caption{Number of HDL Cases ( $10^4$)}
        \label{fig:subfigB}
    \end{subfigure}
    \caption{Comparison of HDL Cases and Bugs Found by Different Optimization Algorithms}
    \label{fig:figure7}
\end{figure}

\subsection{Answer to RQ4}

Considering the indispensable role of the Lin\_UCB algorithm in Lin-Hunter, we conducted an ablation experiment evaluating the Lin\_UCB algorithm component to evaluate its impact on Lin-Hunter's bug detection capabilities. 
Specifically, we compared the efficiency of using Lin\_UCB, $\epsilon$-greedy, and thompson-sampling strategies in selecting CPS model mutation strategies. 
We continuously tested Lin-Hunter for 7 days under the three strategies. The evaluation metrics covered the number of successfully generated HDL test cases and the number of bugs found. 
For the number of successfully generated HDL test cases, we synthesis all generated HDL test cases and record them which can be correctly synthesized. For the number of bugs found, we used the differential testing component to verify the generated HDL test cases, counting the number of bugs found under each strategy. 
This is an ablation study for our Lin\_UCB algorithm.

\textbf{Experiment Results.} 
As depicted in \figurename~\ref{fig:subfigB}, the use of Lin\_UCB achieves the best performance in terms of both the number of HDL test cases generated and the bug detection capability. 
Leveraging the reinforcement learning algorithm Lin\_UCB, Lin-Hunter generated 9.84\% more HDL test cases compared to the thompson-sampling strategy and 28.85\% more than the $\epsilon$-greedy approach. 
These experimental results highlight advantages of the Metamorphic strategy selection Component in improving the overall effectiveness of the framework. 
Guided by the Lin\_UCB algorithm, Lin-Hunter is able to generate more test cases that can be correctly synthesized, thereby demonstrating superior efficiency in bug detection.

\textbf{Summary of RQ4.} 
The Metamorphic strategy selection Component plays a pivotal role in enhancing both the efficiency of HDL test case generation and the subsequent bug detection process. 
By enabling the generation of more complex and diverse test cases, this component significantly improves the capability of Lin-Hunter in identifying challenging corner cases and ensuring effective synthesis.

\section{Conclusions}
\label{sec:conclusion}

In this paper, we propose a novel FPGA logic synthesis tool testing method called Lin-Hunter. 
Our method leverages the equivalence of equivalence mutation strategies to effectively diversify the generation of HDL test cases. 
Furthermore, it employs the Lin\_UCB algorithm with a dynamic reward updating mechanism to guide the selection of mutation strategies based on synthesis log information, thereby increasing the likelihood of triggering previously undiscovered bugs. 
Over a three-month evaluation period, we have reported 16 previously unknown bugs in mainstream FPGA logic synthesis tools, all of which have been independently confirmed by the official developers and 15 of which will be fixed in their upcoming releases. 
In the future, we will explore using large language models (LLMs) to more thoroughly analyze bug reports and investigate more efficient mutation strategies to comprehensively test FPGA logic synthesis tools.

% \begin{acks}

% This work was supported by the National Natural Science Foundation of China (No.62472062), the Dalian Science and technology Innovation Fund project (No.2024JJ12GX022), the Fundamental Research Funds for the Central Universities (No.3132025265).

% \end{acks}

\bibliographystyle{ACM-Reference-Format}
\bibliography{mytosem}

\end{document}